\documentclass[aps,prb,groupedaddress,
                floatfix,
                showpacs,
                amsfonts,amssymb,
               twocolumn,
    ]{revtex4}
\usepackage{amsmath}
\usepackage{graphicx}
\usepackage{dcolumn}
\usepackage{revsymb}
\usepackage{natbib}
\usepackage{array}
\usepackage{xspace}
\usepackage{setspace}
\usepackage{bm}
\usepackage{verbatim}
\newcommand{\Ti}{{T_i}}
\newcommand{\Tf}{{T_o}}
\newcommand{\mui}{{\mu_i}}
\newcommand{\muf}{{\mu_o}}

\newcommand\Ecal{\mathcal{E}}
\newcommand\Dcal{\mathcal{D}}
\newcommand\Tcal{\mathcal{T}}
\newcommand{\relphantom}[1]{\mathrel{\hphantom{#1}}}% for use in split environment
\newcommand{\superlabel}[1]{^{\text{#1}}}%
\newcommand{\nth}[1][n]{\ensuremath{#1\superlabel{th}}\xspace}
\newcolumntype{C}{>{$}c<{$}}
\newcolumntype{L}{>{$}l<{$}}
\newcolumntype{R}{>{$}r<{$}}
\newcolumntype{.}{D{.}{.}{-1}}
\newcolumntype{d}{D{.}{.}{-1}}
\newcommand\Emin{\Ecal_1^i}
\newcommand\Topt{T_{\text{opt}}}
\begin{document}
%
% =============================================================================

\title{Evaporative Cooling in Semiconductor Devices}%

\author{Thushari Jayasekera}
\author{Kieran Mullen}
\author{Michael A.~Morrison}
\email[Electronic address: ]{morrison@mail.nhn.ou.edu}
\affiliation{Department of Physics and Astronomy, The University
of Oklahoma, 440 West Brooks Street, Norman, Oklahoma 73019-0225}
% =============================================================================

\begin{abstract}

We discuss the theory of cooling electrons in solid-state devices via
``evaporative emission.'' Our model is based on filtering electron
subbands in a quantum-wire device. When incident electrons in a
higher-energy subband scatter out of the initial electron
distribution, the system equilibrates to a different chemical
potential and temperature than those of the incident electron
distribution. We show that this re-equilibration can cause
considerable cooling of the system. We discuss how the device
geometry affects the final electron temperatures, and consider
factors relevant to possible experiments. We demonstrate that one can
therefore substantial electron cooling due to quantum effects in a
room-temperature device. The resulting cooled electron population
could be used for photo-detection of optical frequencies
corresponding to thermal energies near room temperature.
\end{abstract}

\pacs{73.50.Lw,73.23.-b}%

\maketitle
% =============================================================================

\section{Introduction
    \label{sec:Introduction}}

As electronic devices become smaller, they leave the regime of
classical physics and enter the realm of quantum physics. Many
classical quantities such as resistance must be reinterpreted for
systems on a mesoscopic scale. One such classical concept is that of
the refrigerator: a device that uses an external source of work to
cool a gas. In this paper we consider whether this classical concept
can be applied to an electron gas so that one could cool such a gas
by applying a voltage to a device.

There are many ways to cool electrons in a condensed-matter system.
For example, thermoelectric coolers based on the Peltier
effect~\cite{peltier} are available commercially.  A different kind
of electron-cooling mechanism in semiconductor devices is based on a
quasistatic expansion of a two-dimensional electron
gas.\cite{kirczenow} Still other possibilities include taking
advantage of many-body effects that can lead to liquid/gas phase
transitions in the electron population in a semiconductor quantum
well.\cite{Xie,unpub}

In this paper we investigate electron cooling in a mesoscopic
solid-state device using evaporative emission. This method entails
removal of (``filtering'') electrons from a high-energy subband of a
many-electron system, followed by relaxation of the remaining
electrons to a temperature lower than that of the initial system.
Evaporative cooling is widely used in bosonic systems.\cite{BEC} But
this method is harder to implement for fermionic systems, as we shall
discuss below.

In section \ref{sec:Theory} we describe the theory of a
two-dimensional device to cool electrons in quantum wires. We use the
Landauer formula\cite{Landauer,ButLand,LandauerReview} to analyze the
cooling properties of these devices. This formula was originally
developed to explain the transport properties of electrons in a
quantum device. It relates these properties to the quantum mechanical
scattering amplitudes for electrons that pass through the device. To
calculate these amplitudes, we use an extension of R-matrix
theory\cite{rmat1,rmat2,rmat3} that we summarize in the Appendix. In
section \ref{sec:Results} we use this theory to calculate cooling
properties of several two-dimensional devices. We begin with a simple
T-junction device and show that by optimizing its design we can
achieve electron cooling. We improve upon this result by switching to
a ``plus-junction'' design, which can give up to 15\%~cooling. In
section \ref{sec:Experiment} we discuss applications and realistic
parameters for a device to cool electrons, and
in~\ref{sec:Conclusions} we summarize our key results and describe
future research.

\section{Electron Cooling in Two-Dimensional Quantum Devices
    \label{sec:Theory}}

Our theoretical approach is analogous to the working principle of the
classical Hilsch vortex tube,~\cite{Hilsch} which uses a T-shaped
assembly of pipes to separate high-pressure air into a
high-temperature system and a low-temperature system. This separation
does not violate the Second Law of Thermodynamics, because the system
is driven by an external force.

\begin{figure}[hbt!]
\centering
\includegraphics[width=3.3 in]{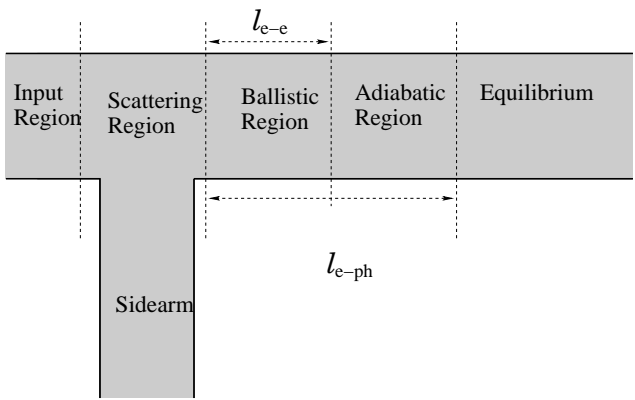}
\caption{Regions in a two-dimensional T-junction device.  The
``ballistic region'' is the part of the device within a distance
comparable to the electron-electron scattering length ($\ell_{e-e}$)
of the junction. The ``adiabatic region'' is within the
electron-phonon scattering length ($\ell_{e-ph}$) of the junction. It
is in the adiabatic region that we achieve cooling.  At larger
distances, in the ``equilibrium region,'' the electrons have returned
to the temperature of the lattice.
    \label{fig:tregions}}
\end{figure}

We use a similar idea to cool electrons in a quantum mechanical
system. The simplest such device uses a T-shaped assembly of quantum
wires to remove higher-energy electrons from an electron gas at fixed
temperature. Figure \ref{fig:tregions} is a schematic of such a
configuration that defines the regions of the device. We assume that
our device is formed from a quantum well whose thickness is
sufficiently small that the device can be considered two-dimensional.
That is, we assume that the device confines electrons to a layer of
thickness $z_0$ such that the confinement energy associated with
motion in the $z$-direction is much larger than any other energy in
the problem. (This confinement energy is~$\hbar^2 \pi^2/2m^*\,z_0^2$,
where $m^*$ is the electron effective mass.)
%We further assume that the width of
%the initial lead, which is denoted by~$w_0$ in~Fig.\ref{fig:tregions}, is
%so small that electrons in this lead can occupy only the first or
%second sub-band.

There are three leads in the T-junction: the input lead, output lead,
and sidearm, as shown in~Fig.\ref{fig:tregions}. We shall label
physical quantities by subscripts ``i,'' ``o,''  and ``s''
respectively; for example,  we denote the widths of the leads
by~$w_i$, $w_o$, and~$w_s$. Electrons are injected into this device
through the input lead, and in the input region are in thermal
equilibrium at an initial temperature $\Ti$ and chemical
potential~$\mui$. Filtering of higher-energy electrons from the
initial electron gas occurs in the \emph{scattering region}. The rate
of scattering into the sidearm depends upon the electron energy: if
the subband energy in lead of width $w$ is given by $E_n=\hbar^2
\pi^2 n^2/2mw^2$, then the ``force'' electrons exert on the sides of
the lead, $F=-dE/dw$, is larger for larger~$n$. Thus when an electron
encounters the sidearm, the higher-subband states are more likely to
squirt down the sidearm. Alternatively, one can see that the
higher-subband electrons scatter preferentially to the sidearm
because they are in states with wave functions that are linear
combinations of plane waves with larger transverse momenta.

Electrons that scatter forward into the output lead proceed into the
\emph{ballistic region}, where the electron population is determined
entirely by the product of the initial electron distribution and the
scattering probability. Next the electrons enter the \emph{adiabatic
region}, where they exchange energy among themselves and so relax to
a temperature~$\Tf$ and a chemical potential~$\muf$. In principle,
the values of these properties can be calculated from conservation
laws for energy and particle number. In section \ref{sec:Results} we
report such calculations and show that for some device geometries the
temperature $\Tf$ is less than the initial temperature: \emph{this
temperature decrease is the desired cooling effect}.

Finally, at large distances (as determined by the electron-phonon
scattering rate) the electrons will return to equilibrium with the
lattice at the initial temperature~$T_i$; this re-equilibration
occurs in the \emph{equilibrium region}. To see cooling, one must
measure the electron temperature before they get to this region

Below we define our notation and describe how we calculate the
electron distributions in the input and output leads. We also define
and describe the calculation of a quantitative measure of electron
cooling in the device.

\subsection{Input electron densities and populations
    \label{subsec:input}}

Provided the electrons in each lead are in thermodynamic equilibrium,
we can treat them as an ideal Fermi gas. For lead~$\ell$, therefore,
the {subband electron density} (per unit volume) in an open
(energetically accessible) subband~$n$~is
\begin{equation}
\label{eq:subbandDensity} \rho^{(\ell)}_n(E,T_\ell,\mu_\ell) =
f(E;T_\ell,\mu_\ell)\,\Dcal^{(\ell)}_n(E)
\end{equation}
for $n$ such that the subband energy obeys $\epsilon_n^{(\ell)} \leq
E$. In this equation, the density of states~is
\begin{equation}\label{eq:dos1D}
    \Dcal^{(\ell)}_n(E)
    = \big[E - \epsilon^{(\ell)}_n\big]^{-1/2}.
\end{equation}
The quantity~$E$ is the dimensionless electron energy measured in
units of the lowest subband energy of the input channel, $\Emin\equiv
\hbar^2\pi^2/2m^*\,w_i^2$. We choose the zero of energy at the energy
of the ground transverse state in the input lead. We measure all
other subband thresholds relative to this energy and in units of
$\Emin$, so that
\begin{equation}
\epsilon^{(\ell)}_n = n^2 \,\dfrac{w_i^2}{w_\ell^2 } - 1.
\end{equation}

The occupation probability for lead~$\ell$ is given by the
Fermi-Dirac distribution function for total electron
energy~$E=E_n^{(\ell)}+\epsilon^{(\ell)}_n$,
\begin{equation}\label{eq:FD}
f(E;T,\mu_\ell) =
    \dfrac{1}{e^{(E-\mu_\ell)/k_B T} + 1},
\end{equation}
where~$\mu_\ell$ is the electrochemical potential in lead~$\ell$,
 $E_n^{(\ell)}$ is the \emph{longitudinal kinetic energy} of the electron,
and $k_B$ is Boltzmann's constant. The corresponding subband
population in lead~$\ell$ is the integral of the subband
density~(Eq.~(\ref{eq:subbandDensity})) over all allowed total
electron energies~$E$:
\begin{equation}\label{eq:subbandpop}
%\subbandpop{n} = \energyint{n} \subbanddensity{n}\dE.
N_n^{(\ell)}(T_\ell,\mu_\ell) =
    \int_{\epsilon_n^{(\ell)}}^\infty\rho_n^{(\ell)}(E;T_\ell,\mu_\ell)\,dE.
\end{equation}
The total population in lead~$\ell$ is the sum of the subband
populations for that lead,
\begin{equation}\label{eq:leadpop}
N^{(\ell)}(T_\ell,\mu_\ell) =
    \sum_{n=1}^\infty N_n^{(\ell)}(T_\ell,\mu_\ell).
\end{equation}

\subsection{Transmitted-electron densities and populations
    \label{subsec:Transmitted}}

Immediately upon leaving the scattering region the transmitted
electrons are in a highly nonequilibrium distribution and cannot be
characterized by a temperature or a chemical potential. By the time
these electrons have traveled a distance along the lead comparable to
several times their relaxation length, they have come to equilibrium
at~$T_o$ and~$\mu_o$, and it is meaningful to describe them by a
Fermi-Dirac distribution function $f(E;T_o,\mu_o)$. These output-lead
properties refer to the electron population in the adiabatic region,
where the electrons are in thermodynamic equilibrium. Given $T_i$ and
$\mu_i$ for the \emph{input} lead, our goal is to determine the
values of $T_o$ and $\mu_o$ for the \emph{output} lead that yields
the lowest $T_o < T_i$; i.e., which maximizes cooling of transmitted
electrons.

For a given~$T_i$ and $\mu_i$, we can determine the temperature in
the output lead~$T_o$ by requiring that the number of electrons in
the output lead at equilibrium equals the number of electrons
transmitted into this lead (conservation of electrons in the output
lead).
%Since experimentalists can
%adjust~$\mu_o$  by changing the bias applied to the electron
%reservoirs that are attached to the input and output leads, we
%simultaneously determine the value of $\mu_o$  that minimizes~$T_o$.
%To this end, we also require conservation of the ensemble-average
%electron energy in the output lead.
To set up equations to implement this strategy, we must define
subband and lead populations in terms of electrons transmitted from
the input lead into the output lead. The (state-to-state) density of
electrons transmitted from an open subband~$n$ of the input lead $i$
into an open subband~$n'$ of a lead~$\ell'$~is
\begin{equation} \label{eq:transDensity}
\rho^{\ell',i}_{n',n}(E;T_i,\mu_i) =
    \rho_n^{(i)}(E,T_i,\mu_i)\,\Tcal^{\ell',i}_{n',n}
\end{equation}
for $E\geq \epsilon^{\rm max}_{n',n}$, where $\Tcal^{\ell',i}_{n',n}
$ is the transmission coefficient from subband~$n$ in lead~$i$  to
subband~$n'$ in lead~$\ell'$. The restriction that the total electron
energy~$E$ be greater than or equal to
\begin{equation}\label{eq:epmax}
\epsilon^{\rm max}_{n',n} \equiv
    \max{\left\{\epsilon_n^{(i)},\epsilon_{n'}^{(\ell')}\right\}}
\end{equation}
ensures that subbands~$n$ and~$n'$ are both open; were this
restriction violated, then the transmission coefficient
$\Tcal^{\ell',i}_{n',n}$ would be undefined.

The population of electrons transmitted into subband~$n'$ of the
output lead, the transmitted subband density,~is
\begin{equation}\label{eq:transmittedSubbandPop}
N^{o,i}_{n'}(T_i,\mu_i) =
    \sum_{n=1}^\infty
    \int_{\epsilon^{\rm max}_{n',n}}^\infty
    \rho^{o,i}_{n',n}(E;T_i,\mu_i)\,dE.
\end{equation}
Hence the total population of electrons transmitted into the output
lead, the transmitted density,~is
\begin{equation}
\label{eq:transmittedLeadPop}
N^{o,i}(T_i,\mu_i) = \sum_{n'=1}^\infty
N^{o,i}_{n'}(T_i,\mu_i).
\end{equation}

\subsection{The cooling parameter
    \label{subsec:Cooling}}

For a given~$T_i$ and $\mu_i$, we can determine the equilibrium
temperature in the output lead~$T_o$ by solving simultaneously the
equation for conservation of the number of electrons,
\begin{equation}
\label{eq:conserveParticles}
N^{(o)}(T_o,\mu_o) = N^{o,i}(T_i,\mu_i),
\end{equation}
and the equation for conservation of electron energy in this lead:
\begin{equation}
\label{eq:conserveEnergy}%
\langle E \rangle^{(o)}(T_o,\mu_o)
    = \langle E \rangle^{o,i}(T_i,\mu_i).
\end{equation}
In Eq.~(\ref{eq:conserveEnergy}) the energy of
 the electrons at equilibrium in the output
lead~is
\begin{equation}
\label{eq:leadEnergy}
\langle E \rangle^{(o)}(T_o,\mu_o) \equiv
    \sum_{n'=1}^\infty \,
\int_{\epsilon_{n'}^{(o)}}^\infty E \, \rho_{n'}^{(o)}(E;T_o,\mu_o) \, dE,
\end{equation}
where the subband density in the output lead,
$\rho_{n'}^{(o)}(E;T_o,\mu_o) $ is defined
by~Eq.~(\ref{eq:subbandDensity}). The energy of the electrons
\emph{transmitted into the output lead}~is
\begin{equation}\label{eq:transLeadEnergy}
\langle E \rangle^{o,i}(T_i,\mu_i) =
    \sum_{n'=1}^\infty \sum_{n=1}^\infty\;
    \int_{\epsilon_{n',n}\superlabel{max}}^\infty
    E\,\rho_{n',n}^{o,i}(E;,T_i,\mu_i)\,dE,
\end{equation}
where the transmitted-electron
density~$\rho_{n',n}^{o,i}(E;,T_i,\mu_i)$ is given
by~Eq.~(\ref{eq:transDensity}).  Using
Eqs.~(\ref{eq:transmittedLeadPop}), and~(\ref{eq:transLeadEnergy}) we
calculate the number of transmitted electrons and their energy. We
then calculate the equilibrium temperature and chemical potential
that would give the same total number and energy. This calculation
produces the parameters~$T_o$ and $\mu_o$ for the output lead.

As a measure of the effectiveness of a given device for cooling
electrons, we define the {cooling parameter}
\begin{equation}\label{eq:coolingParam}
\eta(T_i,\mu_i) \equiv
    \dfrac{T_o(T_i,\mu_i)}{T_i}.
\end{equation}
If~$\eta > 1$, the device \emph{heats} electrons.
Our goal, therefore, is to determine the device geometry and initial
electron properties~$T_i$ and $\mu_i$ that minimize~$\eta < 1$.

%To within a few times~$\kB T_i$, the current from the input lead
%to the output lead is carried by electrons near the Fermi energy~$\eF
%$, where~$\mu_i \leq\eF\leq \mu_o$. The \textbf{optimum voltage
%drop} across the electron reservoirs attached to the input and
%output leads---the voltage required to maximize electron cooling---is
%determined from~$T_i$, $\mu$, and the optimum output-lead
%temperature~$T_o$ as
%\begin{equation}
%e\,V_{\rm opt} = \left[1 - \Tcal^{o,i}_{\rm tot}(E) \right](\mu_i - \mu_o ),
%\end{equation}
%where~$\Tcal^{o,i}_{\rm tot}(E)$ is the total transmission
%function for the output lead.
%

\section{Results
    \label{sec:Results}}

To achieve optimum cooling the higher-subband  electrons should
scatter into the sidearm, and the lower-subband electrons should
scatter into the output  lead.  In a real device not all
higher-subband  electrons will scatter into the sidearm,  and not all
the lower-subband electrons will scatter into the output lead. The
probabilities for electron scattering, as quantified in transmission
coefficients, depend on system properties such as the geometry, and
on scattering potentials. We consider ``perfect devices'' that have
no impurities, so electrons are scattered only by the boundaries of
the device. We further assume that the potential energy of the
electrons in the leads is zero; this simplifying assumption is
\emph{not} essential to either the cooling effect or to our
formalism. We can alter the scattering of electrons by changing the
ratio of the width of the sidearm to that of the input lead
(Fig.~\ref{fig:tregions}) or by changing the geometry altogether.

\subsection{T-junction cooling devices}

To calculate the cooling parameter $\eta$ we need to know the
population of electrons in each subband. To determine this quantity
we must know transmission coefficients from a state in the input lead
to states in the output lead. To calculate these transmission
coefficients we use a generalization of R-matrix theory that we
summarize in the Appendix. To determine the results reported here, we
calculated cooling parameters $\eta$ using transmission coefficients
in a T-junction device for various ratios of the width~$w_s$ of the
sidearm to the width~$w_i$ of the input lead, keeping~the width~$w_o$
of the output lead equal to~$w_i$.

\subsubsection{A T-junction device with $w_s=w_i=1.0$}

\begin{figure}[hbt]
\centering
%\scalebox{0.4}
\includegraphics[scale=0.5]{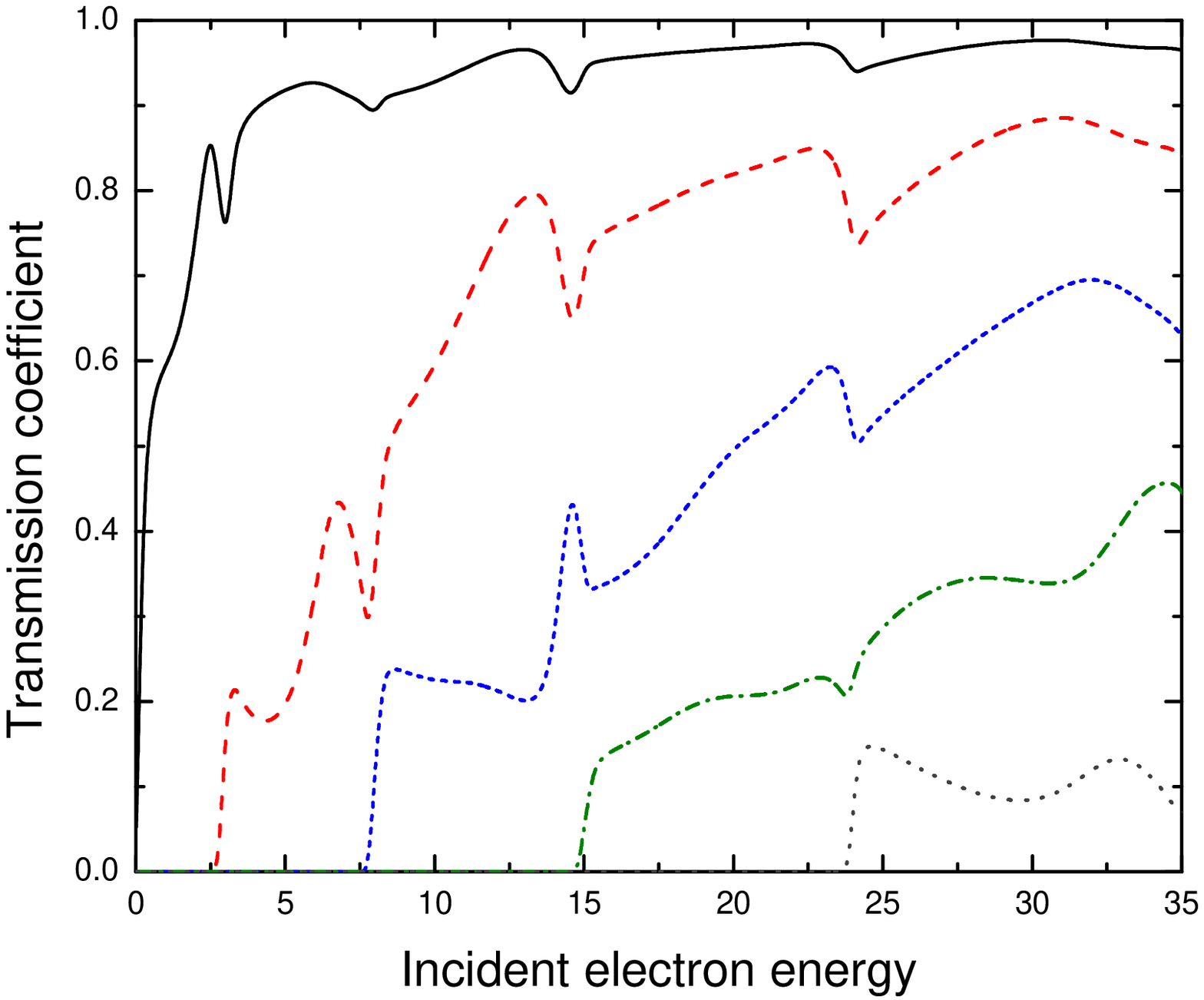}\\
\includegraphics[scale=0.5]{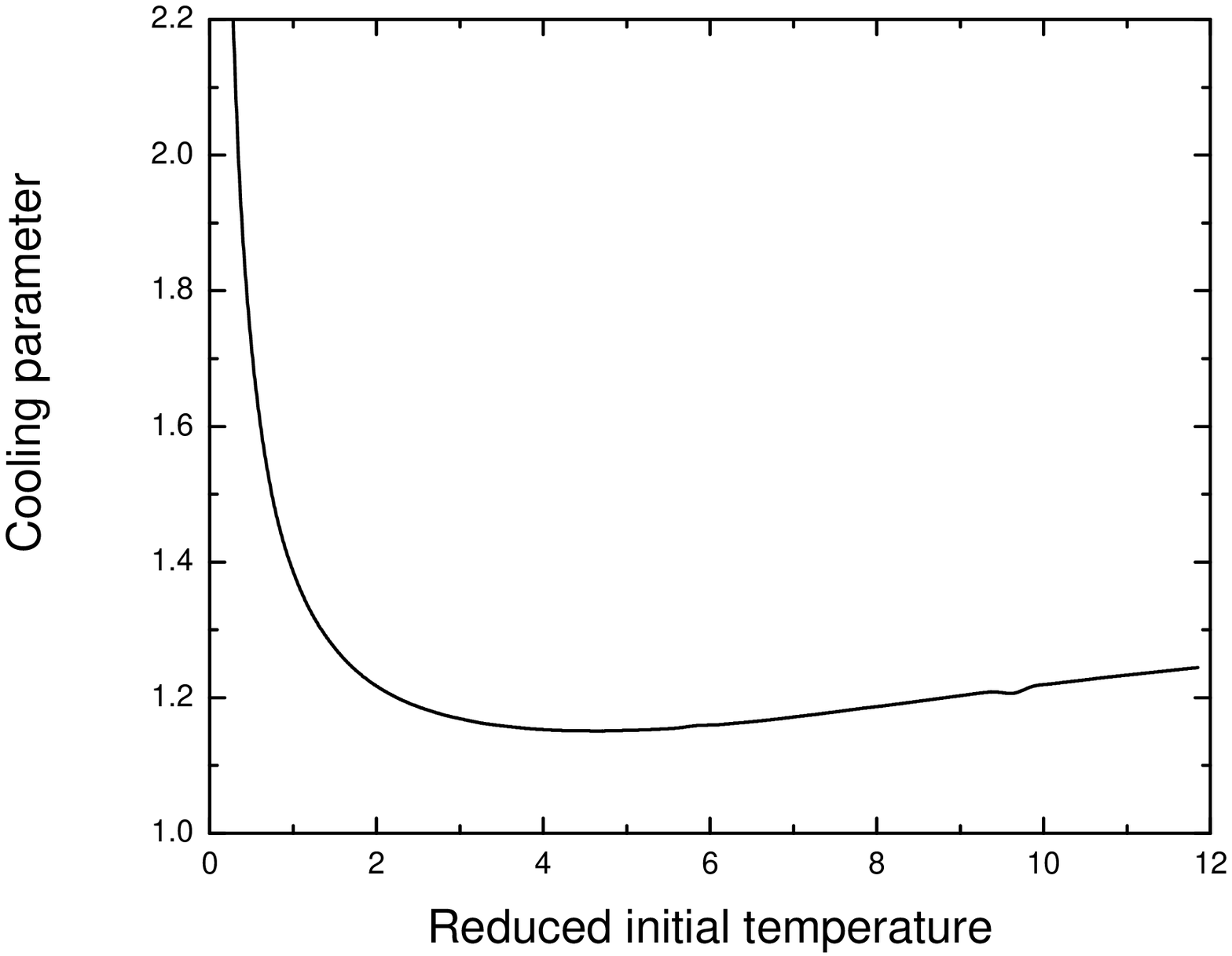}%
\renewcommand{\baselinestretch}{1.2}
\caption{Upper panel: State-to-output-lead transmission coefficients
for electrons in a T-junction device with $w_i=w_s=w_o$. The curves
correspond to different subbands of the incident electrons: $n_i = 1$
(solid curve), 2~(dashed), 3~(short dash), 4~(dash-dot), 5~(dotted).
The horizontal axis is the energy of the incoming electron measured
in terms of the first-subband energy of the input lead,
$\Emin=\hbar^2\pi^2/2m^*\,w_i^2$, from a zero of energy at~$\Emin$.
Lower panel: The cooling parameter $\eta$ for the coefficients in~(a)
for initial chemical potential $\mu_i=0$. The ``reduced initial
temperature'' (horizontal axis) is the dimensionless quantity~$k_B
T_i/\Emin$.
    \label{fig:trans10}}
\renewcommand{\baselinestretch}{1.5}
\end{figure}

We first consider a T-junction device in which all leads have the
same width: $w_i=w_s=w_o$. Figure~\ref{fig:trans10}(a) shows
``state-to-lead'' transmission coefficients for scattering into the
output lead of electrons in different subbands of the input lead.
These coefficients are sums over all energetically accessible
(``open'') subbands~$n_o$ of the output lead of state-to-state
transmission coefficients~$\Tcal^{\ell',i}_{n_o,n_i}$ [see
Eq.~\ref{eq:transDensity}] from a given state~$n_i$ of the input
lead. This figure illustrates the loss of some higher-subband
electrons from the initial electron distribution.

The cooling parameter~$\eta$ for this case is shown in
Fig.~\ref{fig:trans10}(b) for different values of initial
temperature~$T_i$ with the initial chemical potential $\mu_i=0$.
(Note that we measure the energy in terms of $\Emin$, and all the
energies are measured from $\Emin$. So $\mu_i=0$ means that the
external potential of the system is such that the Fermi energy is
$E_F=\hbar^2\pi^2/2m^*\,w_i^2$.) For this geometry~$\eta > 1$ for all
initial temperatures the cooling parameter. That is, this device
\emph{heats} electrons---the opposite of the desired effect.

This case is important because it demonstrates that even if
high-energy electrons are lost due to scattering, a compensatory loss
of low-energy electrons may produce an overall heating effect. Loss
of low-energy electrons opens gaps in the electron distribution at
low energies. Higher-energy electrons can then relax into these newly
accessible low-energy states, with the resulting energy difference
liberated as thermal energy. If this happens, then the loss of
low-energy electrons will \emph{heat} the system. Even a small dip in
the scattered-electron distribution at low energies will
significantly affect the final temperature. To \emph{cool} electrons,
therefore, it is not sufficient to merely scatter higher-energy
electrons. We must scatter thermally excited electrons \emph{but not
significantly scatter
    electrons in lower-energy subbands}.

\begin{figure}[hbt!]
\begin{center}
%\scalebox{0.4}
\includegraphics[scale=0.5]{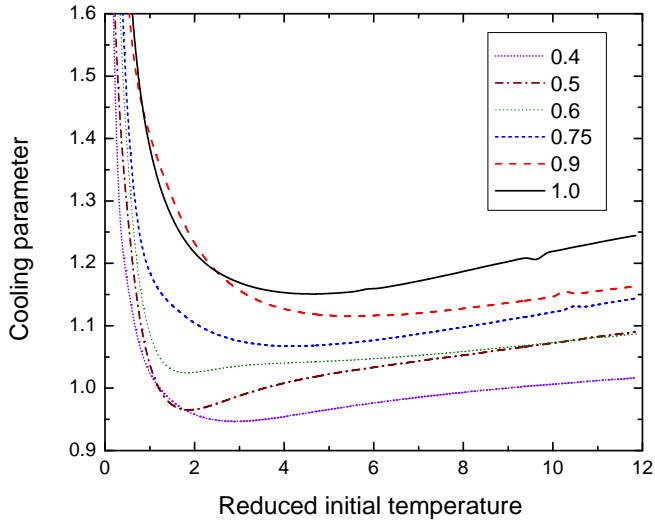}%
\renewcommand{\baselinestretch}{1.2}%
\caption{The cooling parameter $\eta$ as a function of the reduced
initial temperature with initial chemical potential $\mu_i=0$ for
different values of~$w_s/w_i$. The cooling parameter decreases with
increasing magnitude of $w_s/w_i = 1.0$ (solid line), 0.9~(long
dash), 0.75~(medium dash), 0.6~(short dash), 0.5~(dash-dot), and
0.4~(dotted). The reduced initial temperature is defined in the
caption to~Fig.~\ref{fig:trans10}.
    \label{fig:coolingfordifferentT}}
\renewcommand{\baselinestretch}{1.5}
\end{center}
\end{figure}

\subsubsection{Alternative T-junction geometries}

To determine whether a T-junction device can cool electrons, we now
consider several widths~$w_s$ of the sidearm in
Fig.~\ref{fig:tregions}. For each geometry we determine the initial
temperature~$T_i$ and chemical potential~$\mu_i$ that \emph{minimize}
the cooling parameter~$\eta$. Table~\ref{tbl:cooltab} shows these
data and the corresponding final temperature~$T_o$ and chemical
potential~$\mu_o$ for maximum cooling.

\begin{table}[hbt!]
\begin{ruledtabular}
\begin{tabular}{Cddddd}
w_s     & \mu_i     & T_i   & T_o   & \mu_o & \eta \\
\hline
1.0      &   0.0         &   4.65   &   5.35    & -6.24     &     1.15   \\
         &   3.0         &   5.85   &   7.06    &   -5.63   &     1.21    \\
         &   6.0         &   6.77   &   8.69    &   -5.06   &     1.28     \\
\hline
0.9      &   0.0         &   5.68   &   6.34    &   -6.27   &     1.11   \\
         &   3.0         &   6.52   &   7.58    &   -4.97   &     1.16   \\
         &   6.0         &   7.63   &   9.33    &   -4.31   &     1.22   \\
\hline
0.75     &   0.0         &   4.18   &   4.463   &   -3.68   &     1.07   \\
         &   3.0         &   5.36   &   6.01    &   -2.58   &     1.22   \\
         &   6.0         &   6.59   &   7.84    &   -1.80   &     1.19   \\
\hline
0.6      &   0.0         &   1.79   &   1.84    &   -1.09   &     1.02   \\
         &   3.0         &   6.00   &   6.53    &   -1.84   &     1.09   \\
         &   6.0         &   7.39   &   8.40    &   -0.77   &     1.14   \\
\hline
0.5      &   0.0         &   1.85   &   1.79    &   -0.75   &     0.96     \\
         &   3.0         &   3.62   &   3.86    &    0.38   &     1.06    \\
         &   6.0         &   6.52   &   7.35    &    0.58   &     1.13   \\
\hline
0.4      &   0.0         &   2.89   &   2.73    &   -0.71   &     0.95  \\
         &   3.0         &   3.28   &   3.25    &    1.54   &     0.99   \\
         &   6.0         &   5.47   &   5.74    &    2.75   &     1.05   \\
\hline
\end{tabular}
\caption{Cooling parameters~$\eta$ of T-junctions with different
sidearm widths~$w_s$. Also shown are the input chemical
potential~$\mu_i$, output chemical potential~$\mu_o$, and output
temperature~$T_o$ for maximum cooling.
    \label{tbl:cooltab}}
\end{ruledtabular}
\end{table}

To illustrate these data and the effect changing the geometry in this
way, we show in Fig~\ref{fig:coolingfordifferentT} the variation of
$\eta$ with initial temperature~$T_i$ (for initial chemical potential
$\mu_i=0$) for different sidearm widths. These results show that we
achieve cooling ($\eta < 1$) for some geometries and heating ($\eta >
1$) for others. Only the transmission coefficients depend on the
device geometry, so it is through these quantum-mechanical scattering
probabilities that we can control the extent to which a device can
cool electrons.

\begin{figure}[hbt!]
\begin{center}
%\scalebox{0.4}
\includegraphics[scale=0.5]{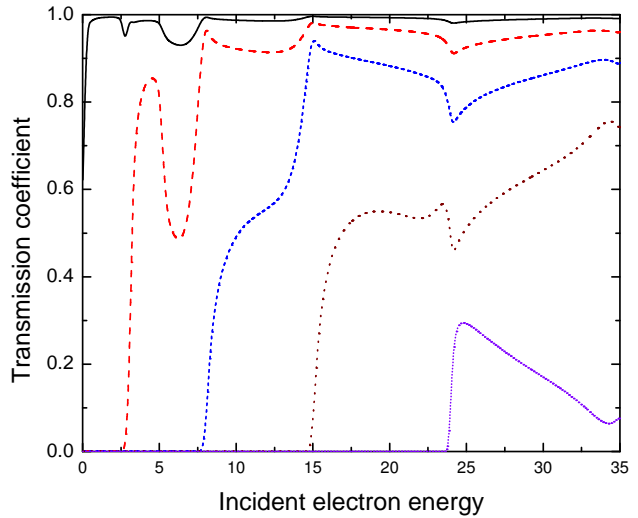}
\renewcommand{\baselinestretch}{1.2}
\caption{State-to-lead transmission coefficients for electrons in a
T-junction device with $w_i=w_o$ and $w_s=0.4 w_i$. Electron are
scattered into the output lead from subbands of the input lead $n_i =
1$ (solid curve), 2~(long dash), 3~(medium dash) 4~(short dash), and
5~(dotted). The horizontal axis is the energy of the incoming
electron measured in terms of the first-subband energy of the input
lead, $\Emin=\hbar^2\pi^2/2m^*\,w_i^2$, from a zero of energy
at~$\Emin$. See also the data in Tbl.~\ref{tbl:cooltab}, which
analyzes cooling for this case.
    \label{fig:0point4trans}}
\renewcommand{\baselinestretch}{1.5}
\end{center}
\end{figure}

We shall now consider in detail a device with $w_s/w_i=0.4$, which
gives $\sim 0.05\%$ cooling. Figure~\ref{fig:0point4trans} shows
state-to-lead transmission coefficients for such a device. Different
curves correspond to different initial subbands.  At low energies the
transmission probability is nearly unity, so for this geometry no
low-energy electrons are lost from the initial distribution. Were we
to adjust the Fermi energy of this device so only the lowest two
subbands were occupied, we would see cooling.

\begin{figure}[hbt!]
\centering%
%\scalebox{0.4}
\includegraphics[scale=0.5]{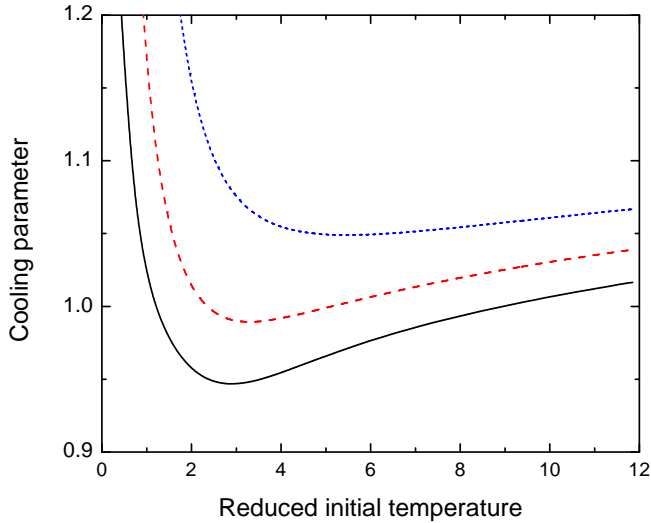}
\renewcommand{\baselinestretch}{1.2}
\caption{The cooling parameter~$\eta$ as a function of the reduced
initial temperature for a T-junction device with $w_i=w_o$ and $w_s =
0.4w_o$ for three initial chemical potentials, $\mu_i=0.0$ (solid
curve), $3.0$ (dashed), and~$6.0$ (dotted). These data are based on
the transmission coefficients shown in Fig.~\ref{fig:0point4trans}.
Maximum cooling is obtained for $\mu_i=0$. The reduced initial
temperature is defined in the caption to~Fig.~\ref{fig:trans10}.
    \label{fig:cooling0point4}}
\renewcommand{\baselinestretch}{1.5}
\end{figure}

In Fig.~\ref{fig:cooling0point4} we illustrate the dependence of the
cooling effect on the initial chemical potential~$\mu_i$. This figure
shows the cooling parameter~$\eta$ for $w_s/w_i=0.4$ as a function of
the initial temperature~$T_i$ for $\mu_i=0.0$, $3.0$, and~$6.0$.
Electron cooling is maximized for~$\mu_i=0$, the edge of the lowest
subband. At this chemical potential all electrons scattered into the
sidearm are in the thermally active region of the Fermi distribution.
Cooling is also obtained for $\mu_i=3.0$, the edge of the second
subband.  But at larger values of $\mu_i$ the device heats electrons.

This example shows that a T-junction \emph{can} cool electrons. We
would prefer, however, a device that produces more cooling
than~$0.05\%$. Investigation of other T-junction geometries showed
that such a device cannot produce significantly more cooling for any
value of~$w_s/w_i$. So we next investigate the addition of a second
sidearm. This change produces the ``plus junction'' illustrated
in~Fig.~\ref{fig:plusfig}.

\subsection{A ``plus-junction'' cooling device}

\begin{figure}[hbt!]
\centering
\includegraphics[width=3. in]{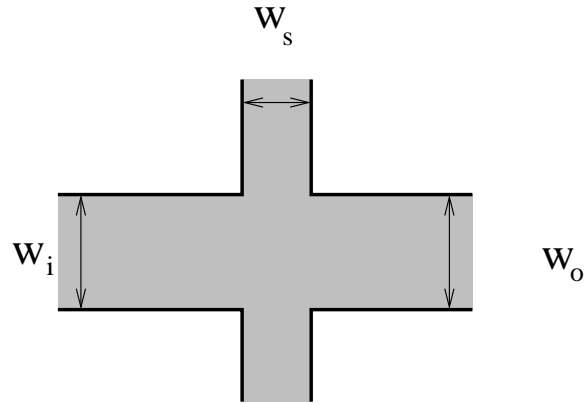}
\caption{Schematic of a plus-junction.  Note that while the widths of
the input and output leads are equal ($w_i=w_o$), the width~$w_s$ of
the two sidearms may be smaller or larger than~$w_i$. This geometry
gives substantially more cooling than any device we have explored
that has only a single sidearm.
    \label{fig:plusfig}}
\end{figure}

\begin{figure}[htb!]
\centering
%\scalebox{0.4}
\includegraphics[scale=0.5]{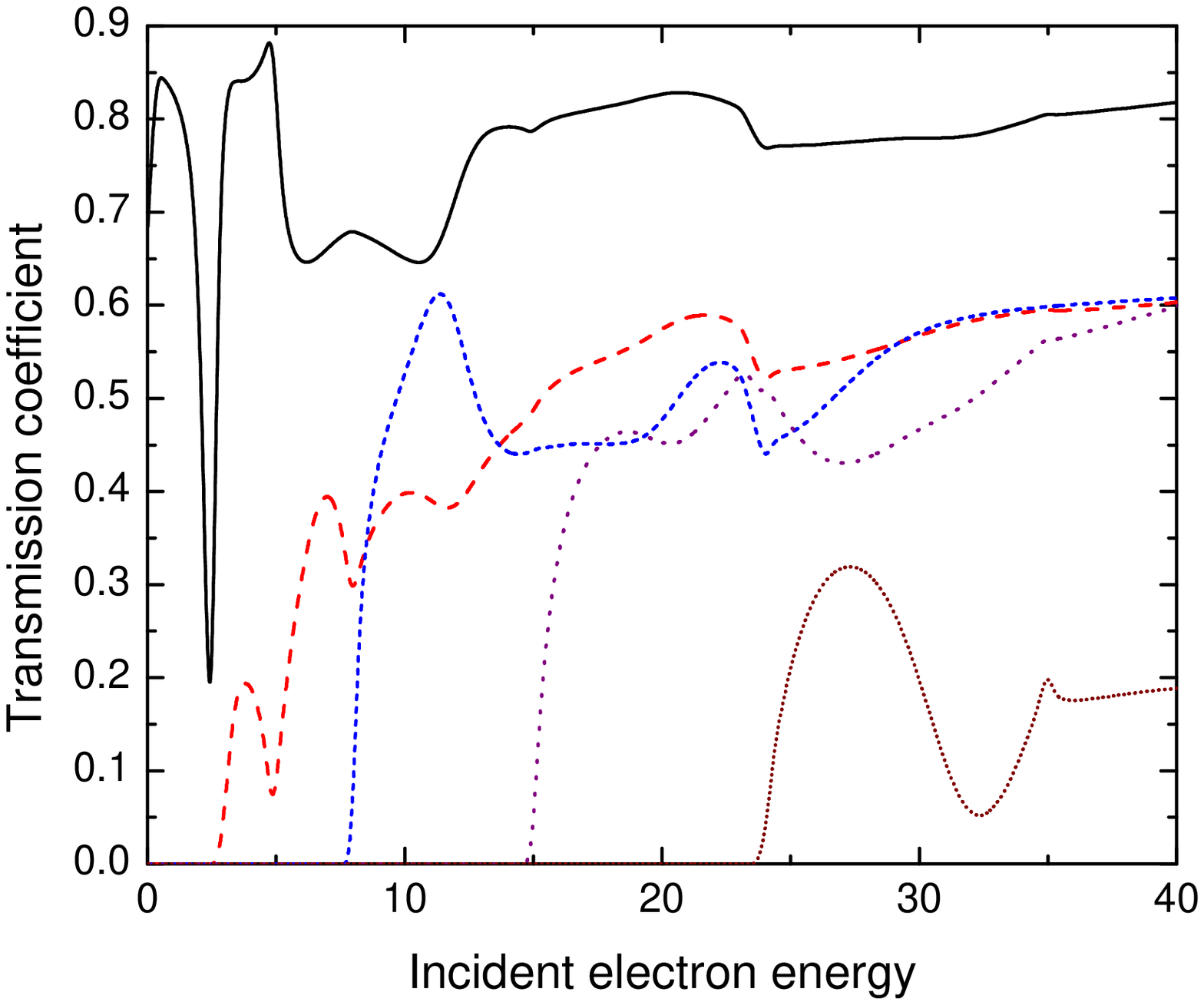}\\
\includegraphics[scale=0.5]{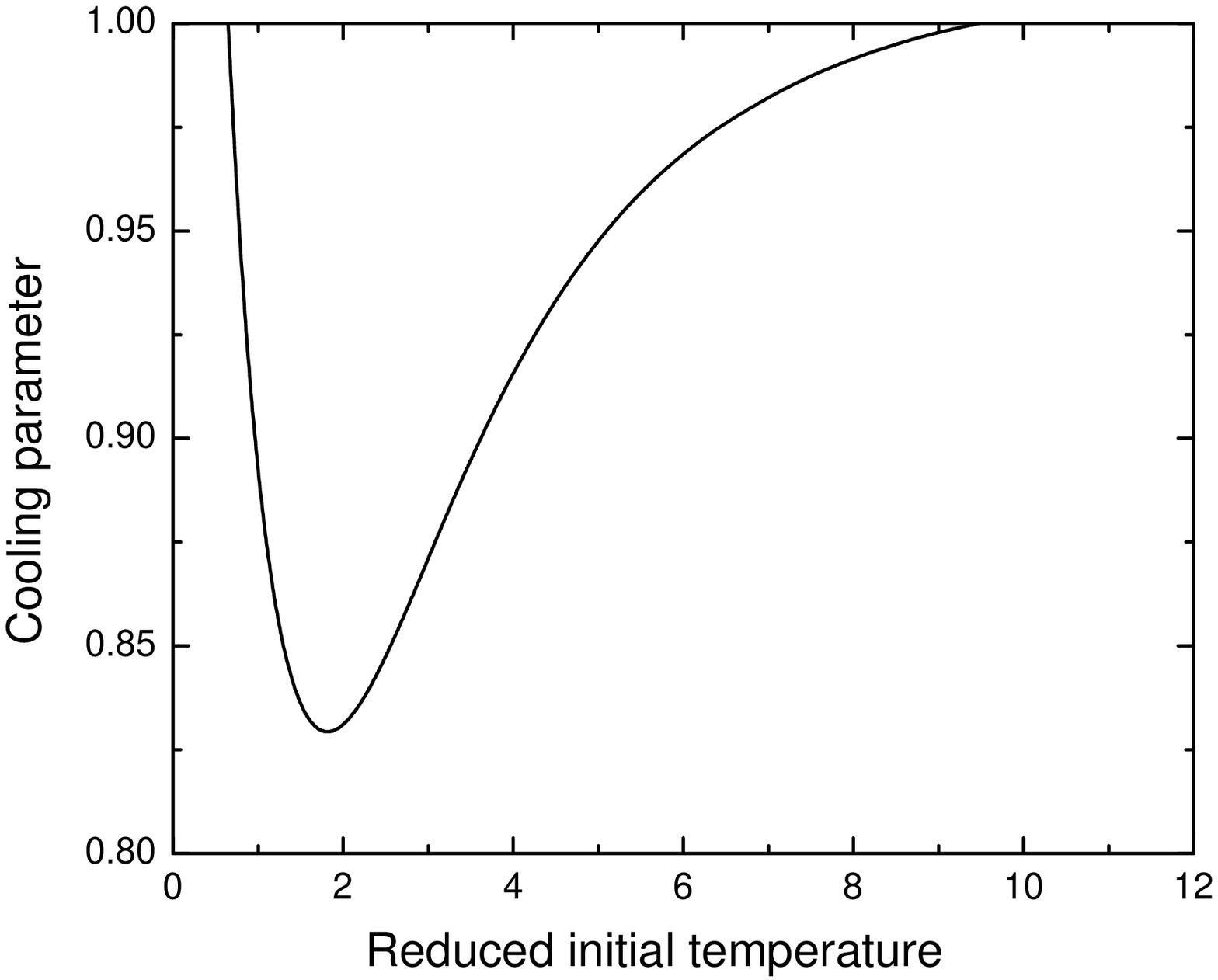}%
\renewcommand{\baselinestretch}{1.2}
\caption{Upper panel: state-to-output-lead transmission coefficients
for electrons in a plus-junction device with $w_s=0.4w_i$. The curves
correspond to different subbands of the incident electrons: $n_i = 1$
(solid curve), 2~(dashed), 3~(short dash), 4~(dash-dot), 5~(dotted).
The horizontal axis is the energy of the incoming electron measured
in terms of the first-subband energy of the input lead,
$\Emin=\hbar^2\pi^2/2m^*\,w_i^2$, from a zero of energy at~$\Emin$.
Lower panel: Cooling parameter $\eta$ based on the coefficients shown
in the upper panel. The reduced initial temperature is defined in the
caption to~Fig.~\ref{fig:trans10}.
    \label{fig:coolplus}}
\renewcommand{\baselinestretch}{1.5}
\end{figure}

\begin{figure}[htb!]
\centering%
%\scalebox{0.4}
\includegraphics[scale=0.5]{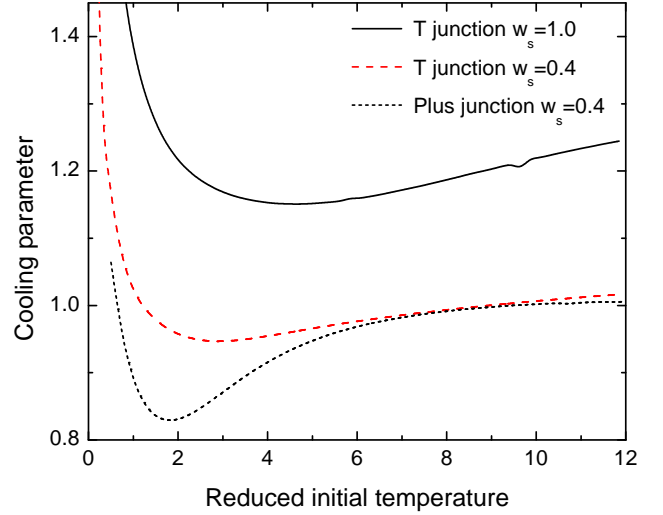}
\renewcommand{\baselinestretch}{1.2}
\caption{Cooling parameters for a T-junction with $w_i=w_s=w_o$
(solid line), a T-junction with $w_i=w_o$ and $w_s/w_0=0.4$ (dashed),
and a plus-junction with $w_i=w_o$ and $w_s=0.4w_i$ (short dash). In
all cases the initial chemical potential is~$\mu_i=0$. The reduced
initial temperature is defined in the caption
to~Fig.~\ref{fig:trans10}.
    \label{fig:cooltogether}}
\renewcommand{\baselinestretch}{1.5}
\end{figure}

Since we achieved cooling in a T-junction device with $w_s=0.4$, we
shall consider a plus junction with the same lead ratios: $w_i=w_o$
and $w_s/w_o=0.4$. (Note that the widths of the two sidearms
in~Fig.~\ref{fig:plusfig} are the same.) State-to-lead transmission
coefficients for this device are shown in Fig.~\ref{fig:coolplus}(a),
and the resulting cooling parameter~$\eta$ as a function of initial
temperature~$T_i$ in Fig.~\ref{fig:coolplus}(b). The latter figure
shows that this geometry yields appreciably more electron
cooling---greater than $15\%$---than did the T-junction devices
of~Tbl.~\ref{tbl:cooltab}. A single unit of a plus-junction device
with this geometry would cool room-temperature electrons by about
$\Delta T=-45C$. Cooling could be further enhanced by combining
several such units in sequence. Comparison of this plus-junction to
the T-junctions discussion previously, as illustrated in
Fig.~\ref{fig:cooltogether}, show how greatly adding a sidearm
improves the effectiveness with which this simple device cools
electrons.

One could try to further optimize the geometry of this device by, for
example, increasing the width of the sidearms with increasing
distance from the junction. Alternatively, one could round the sharp
corners at each junction into smooth curves. However, our initial
explorations of such alterations (for a T-junction) did not produce
substantially more cooling than we obtained with the far simpler plus
junction in~Fig.~~\ref{fig:plusfig}. The essential features that
makes the plus junction more effective than any T~junction is the
presence of more than one sidearm. It is also essential the threshold
for  scattering  into  these sidearms lie above the threshold for the
second subband of the input lead.

\section{Experimental considerations
    \label{sec:Experiment}}

We now consider a plus-junction device using realistic experimental
parameters for Insb and GaAs.  While in Sec.~\ref{sec:Results} we
reported results in dimensionless units, we here use dimensional
units.

All material properties depend on the Fermi energy~$E_F$ of the
system. This quantity is inversely proportional to the effective
mass~$m^*$ of the electrons in the material and is determined by the
electron density in the reservoir. For our device we chose for the
initial chemical potential (which is approximately equal to the Fermi
energy) $\mu_i=0$. Since we have scaled the energy by $\Emin$ and
chosen~$\Emin$ as the zero of energy, setting the initial chemical
potential equals zero means that
\begin{equation}\label{determintthewidth}
\mu_i=E_F= \frac{\hbar^2 \pi^2}{2 m^*\, w_i^2},
\end{equation}
where $w_i$ is the width of the input lead of the device in
Fig.~\ref{fig:plusfig}.  The condition~(\ref{determintthewidth})
allows us to determine the width of the quantum wire as a function of
electron density. Since the Fermi energy and the subband energies
depend on the effective mass in the same fashion, the width of the
quantum wire is independent of the material. For a sample with
electron density~$n$, we have
\begin{equation}
\frac{\pi\hbar^2}{m^*}\,n =
    \frac{\hbar^2\pi^2}{2m^*\,w_i^2},
\end{equation}
which gives for the width of our quantum wire
\begin{equation}
w_i = \sqrt{\frac{\pi}{2\,n}}.
\end{equation}
If, for example,  $n=1.0 \times 10^{11}\,\text{cm}^{-2}$, we obtain
$w_i=39.6\,\text{nm}$, quite a small value. We can increase this
value by decreasing the electron density.

For a plus-junction, we were able to maximize cooling by setting the
initial temperature to $T_i \sim 2$ in dimensionless units. In
dimensional units, this optimum temperature is
\begin{subequations}
\begin{equation}
k_B T = \Topt\,\Emin,
\end{equation}
Using Eq.~(\ref{determintthewidth}), we obtain
\begin{equation}
k_B T=\Topt\, E_F.
\end{equation}
\end{subequations}

For a sample with $n=1.0\times 10^{11}\,\text{cm}^{-2}$, the initial
temperature for maximum cooling is $T_i \sim 82 K$ for GaAs and
$T_i\sim 399 K$ for InSb. Room temperature (300K) corresponds to
\mbox{$T\sim 1.5$} for InSb. \emph{One can therefore obtain
substantial cooling due to quantum effects in a room-temperature
device.} A cooling parameter of $\eta \sim 0.9$ implies that the
electron population is cooled by~$30K$. The resulting cooled
electrons could be used for photo-detection of optical frequencies
corresponding to thermal energies near room temperature.

The devices we have considered have only one cooling stage. One could
increase cooling by connecting multiple plus junctions in series. The
spacing between junctions, however, must be large enough that
resonances in scattering between sidearms are negligible. If not, one
would have to treat the device as a single large quantum mechanical
scattering target. While the presence of such resonances would not
preclude cooling, it would make calculations for a chain of junction
devices more difficult and sensitive to details of phase breaking.

\section{Conclusions and prospects for future research
    \label{sec:Conclusions}}

Many photo-detection applications require a cold detector. We have
presented results for a prototype device that demonstrates electron
cooling in a single-particle picture. We have shown that, while a
naive T-junction can produce modest cooling and may produce heating,
adding an additional sidearm yields a device that can produce
appreciable cooling ---at least~$15\%$. The abrupt discontinuities in
the confining potentials in these models are not essential to
cooling; what is essential is that higher-subband states, which
consist of states with larger transverse momenta, scatter appreciable
into the sidearms. We therefore expect electron cooling in such
devices to be insensitive to details of the potential so long as the
potential does not eliminate the states of the lowest subband.

% -----------------------------------------------------------------------------

\begin{acknowledgments}
This project was supported in part by the US National Science
Foundation under Grant~\mbox{MRSEC DMR-0080054}, and
\mbox{EPS-9720651}, and~\mbox{PHY--0071031}. One of the authors
(TJ)  would like to acknowledge the University of Oklahoma,
Graduate College for travel and research grants.
\end{acknowledgments}

% -----------------------------------------------------------------------------

\begin{appendix}

\section{R-matrix theory for a 2-D system}

\begin{figure}[hbt!]
\centering%
\includegraphics[width=3.3 in]{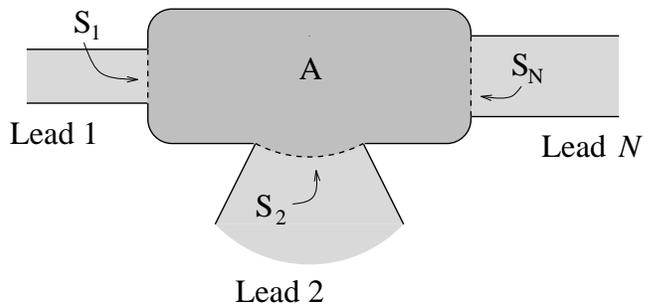}
\caption{Schematic of a two-dimensional device for the present
scattering calculations. The surfaces~$S_1$, $S_2$, \ldots $S_N$
separate the interior region~$A$ from the $N$~leads.
    \label{fig:2dfig}}
\end{figure}

We consider the two-dimensional system in~Fig.~\ref{fig:2dfig}. This
system has a central region~$A$ connected to $N$~external regions or
``leads.'' The leads and the interior region meet at a set of
boundary surfaces we denote by~$S_0$, $S_1$, \dots, $S_N$. We treat
the boundaries between the shaded and unshaded regions as ``hard
walls'' (infinite potential) so electron wave functions are non-zero
only in the shaded regions. Since there may be more than three leads,
we depart from the notation used in the body of this paper (in which
the input, output and sidearm leads were denoted by subscripts $i$,
$o$, and $s$) and denote the input lead by a zero subscript and all
other leads by positive integer subscripts.  We measure all distances
in units of $w_0$ and energies in terms of $E_0\equiv
\hbar^2/2m^*\,w_0^2$. We seek an analytic solution for the amplitudes
of outgoing states in the leads when only one incoming state is
occupied.

The time-independent Schroedinger equation for the scattering
function is
\begin{equation}
\left( \hat{H}-E \right)
    \left|\Psi_{E,n_o}\right\rangle =0.
\end{equation}
where $\left|\Psi_{E,n_0}\right\rangle $ represents the state of an
electron with kinetic energy~$E$ incident in input-lead
subband~$n_0$. Note that $\left|\Psi_{E,n_0}\right\rangle$ is
well-defined in all leads. In a \emph{finite} region, the
Hamiltonian~$\hat{H}$ is not Hermitian. We can produce a Hermitian
operator by adding to $\hat{H}$ the so-called Bloch
operator~$\hat{L}_B$.\cite{rmat2} We denote the eigenfunctions of the
sum of these operators in the interior region by
$\left|\phi_i\right\rangle$ and write the so-called Bloch eigenvalue
equation~as
\begin{equation}
\left( \hat{H}+\hat L_{B}\right) \left| \phi_i\right\rangle= E_i
\left| \phi_i\right\rangle.
\end{equation}
Inserting the Bloch operator into the Schroedinger equation we get
\begin{equation}
\left( \hat{H}+\hat L_{B}-E \right) \Psi_E= \hat L_{B} \Psi_E.
\end{equation}

We now expand the scattering wave function
$\left|\Psi_E\right\rangle$ in the set of orthonormal Bloch
eigenfunctions
%$ (\left| \phi_j \right\rangle \in  D(H_{B})$ in the interior region,
\begin{equation}
\left|\Psi_E \right\rangle = \sum_j C_j\left| \phi_j \right\rangle.
\end{equation}
Inserting this expansion into the Schroedinger equation and using the
properties of the Bloch eigenfunctions yields
\begin{equation}
\left| \Psi_{E,n_0} \right\rangle =\sum_j \frac{\left< \phi_j \right|
\hat L_{B} \left| \Psi_{E,n_0}\right\rangle }{E_j-E} \left| \phi_j
\right\rangle,
\end{equation}
where $E_j$~is the eigenvalue corresponds to the Bloch eigenfunction
$\left| \phi_j \right\rangle$. This expansion is valid throughout the
interior region~$A$ and on its surface (see Fig.~\ref{fig:2dfig}).

To derive an equation for the R~matrix, we now apply this expansion
of the scattering state on each boundary $S_i$. At each such boundary
we can expand the scattering function in either lead eigenfunctions
or Bloch eigenfunctions in the interior region. To be specific, we
introduce a local Cartesian coordinate system for each lead: $x_q$
and $y_q$ are the longitudinal and transverse coordinates of the
\nth[q] lead, respectively. We choose $x_q=0$ on each boundary. (One
can easily choose any orthonormal coordinate system, \emph{mutatis
mutandis}). Each lead eigenfunction is then a product of a plane wave
in the $x_q$ direction and a transverse bound-state
eigenfunction~$\chi_n(y_q)$. The scattering wave function in
the~\nth[q] lead therefore becomes
\begin{equation}
\begin{split}
\Psi_{E,n_0}(x_p,y_p) &=
    e^{-i k_{0,n_0}\, x_0}\,\chi_{0,n_0}(y_0)\, \delta_{p,0}\\
    &\relphantom{=}{}+\sum_{q,n_q=1}^N
    \tau_{q,n_q}(E) e^{i k_{q,n_q}\,x_q}\,\chi_{q,n_q}(y_q)\,\delta_{p,q},
\end{split}
\end{equation}
where $k_{q,n_q}$ and $\tau_{q,n_q}$ are the wave vector and
transmission \emph{amplitude} for the channel with quantum
number~$n_q$ in channel~$q$. Also, $\chi_{n_q}(y_q)$ is the \nth[n_q]
transverse eigenfunction of lead~$q$. Finally, $\delta_{p,q}$ is the
Kroniker delta-function, which ensures that each wave function is
defined only in one lead. If we measure energy in units of $E_0$ then
we can express energy conservation in lead $q$ as $E=k^2_{q,n_q}+
{n^2_q\pi^2}/{w^2_q}$, where $w_q$ is the width of the \nth[q] lead
(in units of~$w_0$). We use this equation to determine the wave
vector $k_{q,n_q}$.

After some algebra we get a set of linear algebraic
equations that we can solve for the transmission amplitudes:
\begin{equation}
\begin{split}
&i \sum_{p,n_p} \tau_{p,n_p}(E) \, k_{p,n_p}
M_{q,n_q,p,n_p}(E)-\tau_{q,n_q}(E) \\
&\relphantom{=}{}= \delta_{q,0}\delta_{n_q,n_0}+i k_{0,n_0}
M_{q,n_q,0,n_0}.
\end{split}
\end{equation}
In writing these equations we have defined
\begin{equation}
M_{q,n_q,p,n_p} =
\int_{y_p}\int_{y_q}\chi^*_{q,n_q}(y_q)R_E(y_q,y_p)\chi_{p,n_p}(y_p)\,dy_q\,dy_p.
\end{equation}
Finally, the R-matrix is given by
\begin{equation}\label{eq:Rmatrix}
R(E,y_p,y_q)\equiv
    \sum_j\dfrac{\phi_j^*(x_q=0,y_q)\, \phi_j(x_p=0,y_p)}{E_j -E}.
\end{equation}
This equation is general in that we can easily adapt it to any number
of leads and to different choices of input~lead.

\end{appendix}
% -----------------------------------------------------------------------------

%\clearpage
\bibliographystyle{apsrev}
%\bibliography{/biblios/physjabb,/biblios/NO,/biblios/books,/biblios/misc,newrefs,BEC07172003,/biblios/Ultracold}

\clearpage
% -----------------------------------------------------------------------------

%\newpage
%\printtables % remove when remove endfloats option
%\newpage
%\printfigures % remove when remove endfloats option

%
\end{document}